# Spatial data sharing with secure multi-party computation for exploratory spatial data analysis

Shuo Shen [1,4], Xinyan Zhu[1,2,3,*], Yanlei Ma [1], XIe Xiang [5], Sun Lilin [5], Xie Hongjun [5], An Rui [5]

1  State Key Laboratory of Information Engineering in Surveying, Mapping and Remote Sensing (LIESMARS), Wuhan University, Wuhan 430079, China; shenshuo@whu.edu.cn
2  Collaborative Innovation Center of Geospatial Technology, Wuhan 430079, China; xinyanzhu@whu.edu.cn
3  Key Laboratory of Aerospace Information Security and Trusted Computing of the Ministry of Education, Wuhan University, Wuhan 430079, China; xinyanzhu@whu.edu.cn
4  Aerospace and Geodesy, Technical University Munich, Munich 80333, Germany; shenshuo@whu.edu.cn
5  Juzix Technology Co., Ltd, Shenzhen 518000, China
*  Correspondence: xinyanzhu@whu.edu.cn; Tel.: +86-13808691386

**Abstract:** Spatial data sharing plays a significant role in opening data research and promoting government agency transparency. However, valuable spatial data, like high-precision geographic information and personal traffic records, cannot be made public because they may incur leakage risks such as intrusion, theft, and the unauthorised sale of proprietary information. When participants with confidential data distrust each other but want to use the other datasets for calculations, the most common solution is to provide their original data to a trusted third party. However, the trusted third party frequently risks being attacked and having the data leaked. To maintain data controllability, most companies and organisations refuse to share their data. In this study, we introduce secure multi-party computation to spatial data sharing to address the sharing problem. Additionally, we describe the design and implementation of the protocols of two exploratory spatial data analyses: global Moran's I and local Moran's I. Furthermore, we build a system to demonstrate process realisation and results visualisation. Comparing our system with existing data-sharing schemes, our system Identifies the correct result without incurring leaking risks during spatial data sharing.

## 1. Introduction

With the rapid development of geographic information, spatial analysis has played a significant role in many fields, such as medicine [1, 2], tourism [3], and sociology [4]. Exploratory spatial data analysis (ESDA), the most representative type of spatial analysis, has especially seen a surge in its use'; it helps solve many problems, like economic [5, 6], mineral substance [7], delivery [8], and health problems [9, 10]. ESDA can even be utilised to judge image quality [11] and street service [12].

All of these significant contributions are related to spatial data. However, data owners can be reluctant to share, especially when the spatial data involves proprietary information, because sharing is accompanied by risks such as the leakage and proliferation of valuable information which can threaten security [13], which increases sharing difficulties. For example, there are two participants, an online shopping enterprise and health bureau, who want to know the distribution relationship between shopping habits and illness. Because the information is too sensitive to share, these organisations do not want to release their original data. How can these institutions learn about this relationship without learning about the attributes of the other dataset?

This scenario is extremely common. On the one hand, geographic information is universally demanded by society. To improve production and livelihoods, we need to public more abundant, large-scale, high-precision, multi-temporal, and multi-factor geographic information, and it is necessary to share professional production data between departments. On the other hand, some data are related to the national economy and individual livelihoods, which must be kept confidential, and users are not allowed to use or spread this information without authorisation. This conflict impedes the development of the geoinformation industry, scientific research, and information infrastructures [14].

Data are building blocks of science. As spatial science grows more data-intensive and collaborative, data sharing becomes more important. Data sharing restrictions therefore impede societal and geoinformational development. In this study, we attempt to dispel the misgivings of holders by performing calculations without risking data leaks. The challenge here is to build a protocol for spatial computation that ensures data security. To accomplish this goal, we introduce secure multi-party computation (SMPC) to spatial data analysis and utilise typical spatial statistical methods, global Moran's I and local Moran's I, which are exploratory spatial data analysis (ESDA) methods, as examples. ESDA, a collaborative task based on spatial data sharing technology, is a suitable representative of the application [15].

The related content is organised as follows. Section 2 describes the research setting of data sharing; Section 3 accounts for the SMPC system running principle and working process, including each service function and SMPC construction; and Section 4 uses global Moran's I and local Moran's I as examples to discuss how to combine SMPC with ESDA. During this process, the formulas are disassembled into three parts to realise respectively. Additionally, we constructed a test platform and compared the time efficiency and accuracy of the results for the traditional and proposed platform, which demonstrates that the proposed platform has practical value.

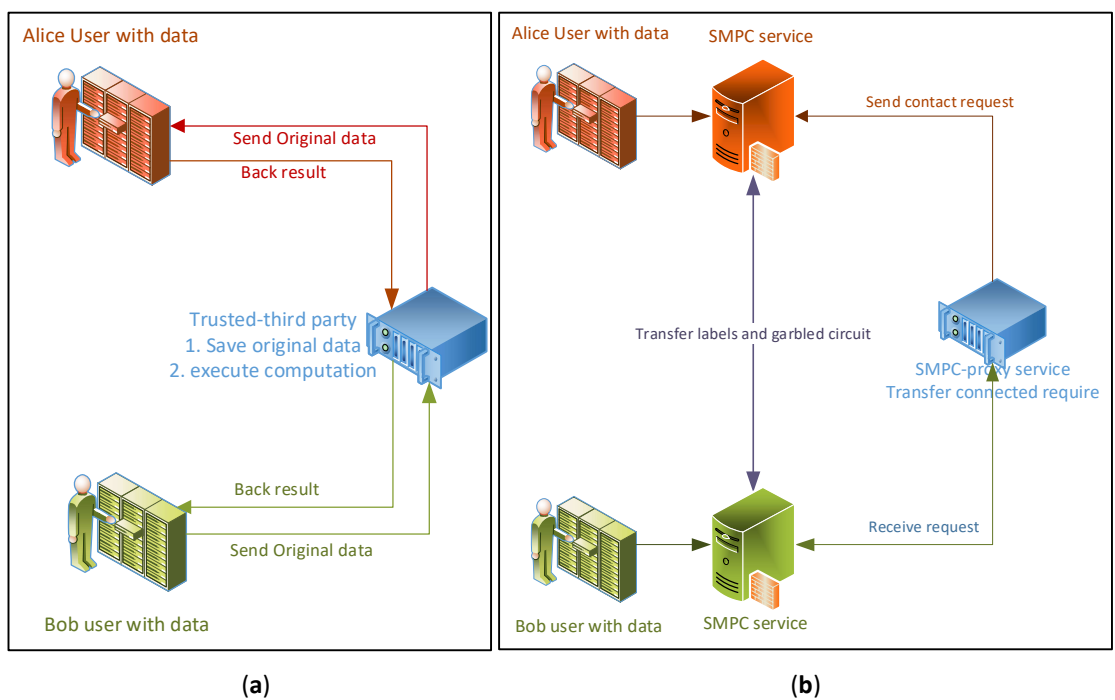

**Figure 1.** This figure shows the calculation structure for mixed data sources. (**a**) Description of Traditional process. (**b**) Description of Security multi-party process

## 2. Literature Review

Identifying and reducing security risks when integrating mixed resources is critical for reassuring data providers. Many studies fuse knowledge from various datasets for commercial activities or scientific research using a trusted third party or encryption methods[16].Introducing a trust third party who is responsible for protecting data and calculating the desired algorithm is a common data integration method. The traditional method is constructed similar to Figure. 1 (a). The users must send the original data to the trusted third party, who then uses the original data to calculate the result. This process can occur in the cloud [17-19] or locally [20, 21], even with limited computing power [22]. Data-market is one of the typical representation, which provide cloud storage [23] or cloud computing [24]. However, in this method, users should submit the original data. And no matter how secure the third party is, there is still the risk of single node invalidation and hacking. Once the central database is paralyzed, the entire service is impossible to fulfil.

Many researchers currently use encryption methods to protect data. However, most encryptions are centred around data stream security. Hur developed a novel ciphertext-policy attributed-based encryption scheme for a data-sharing system by exploiting the system architecture characteristics [25]. The scheme asks clients who own data to upload the data into the external data-storing centre for easier sharing. Pei [26] proposes an identity- and attribute-based signcryption algorithm for enhancing the security of stored and shared remote data. Zhang [27] make great contribution on secure data sharing in smart grid. The scheme uses the technique of inner product encryption, where attribute values are not sent with the shared data. Moreover, some researchers [28, 29] introduce blockchain technology to data sharing. However, the blockchain efficiency in this application is limited. To tackle this problem, Dong Zheng [12] introduce online/offline encryption technology in the encryption phase. However, this requires the attribute centre to be completely trustworthy, which is like a trust third-party.

Because spatial data sharing problems have recently received greater solicitude, we wanted to remove these risks from third parties. SMPC is the most representative solution, but there still a gap regarding combining spatial data and SMPC. Therefore, we use ESDA as an example to build a new structure to demonstrate the integration of multiple spatial resources and execution of spatial data analysis using SMPC.

## 3. SMPC Architecture for ESDA

### 3.1 SMPC concepts

Yao introduced the garbled circuit (GC), which has been used to construct a secure two-party computation framework [30]. The GC can prevent data leaks and obtain accurate results for solving the millionaires' problems [31] related to performing computations with different datasets and without a trusted third party. This framework belongs to the cryptography field and allows two distrustful participators to obtain correct function results without transmitting any original data. The prototype is the basis of the (SMPC). However, there are still many problems will this framework as its protocol can only allow two members to participate at once. Goldrcich perfected the multi-party algorithm and put forward effective porting methods [32]. Time efficiency disadvantages were later addressed [33]. This work has been improved upon, but has not yet been applied to solve practical problems. A growing number of techniques have been used in the real world [34]such as analysing financial data [35], identifying common contacts on smartphones [36], and genome-wide association [37].

There are two main threat models for SMPC. One is the semi-honest (passive) adversary who follows the prescribed protocol but attempts to learn more information than is allowed according to the protocol transcript. Another threat is the malicious (active) adversary who can run an efficient strategy to carry out their attack [38]. In this study, we use the GC-based SMPC framework under the semi-honest model. During the entire process, the parties involved do not worry about the data leakage risks of using the framework when they encounter passive adversaries.

To understand how this works, let us begin by assuming that there are two parties, $P_\alpha$ and $P_\beta$, deployed by a Boolean circuit that only has an or-gate. There are three or-gate wires on the Boolean circuit: two input wires, including $\alpha$ and $\beta$, and one output wire $\gamma$. $P_\alpha$ inputs the data $b_\alpha$ into wire $\alpha$, and $P_\beta$ enters $b_\beta$ into wire $\beta$. The aim is to obtain the result of the gate without leaking any data. Therefore, in the beginning, $P_\alpha$ must generate six random garbled strings, $K_\alpha^0$, $K_\alpha^1$, $K_\beta^0$, $K_\beta^1$, $K_\gamma^0$, and $K_\gamma^1$, as the secret keys to replace the 0 and 1 values of each wire. The functions of these keys are as follows:

$$K_\gamma^0 = K_\alpha^0 \text{ or } K_\beta^0$$
$$K_\gamma^1 = K_\alpha^0 \text{ or } K_\beta^1$$
$$K_\gamma^1 = K_\alpha^1 \text{ or } K_\beta^0$$
$$K_\gamma^1 = K_\alpha^1 \text{ or } K_\beta^1$$

Then, the function we want to calculate can be expressed as: $\mathrm{E}^s_{K_\alpha^{b_\alpha} K_\beta^{b_\beta}}(m)$,

where m is the output wire value ($K_\gamma^0$, $K_\gamma^1$), and S is additional information that is unique to each key to differentiate each choice.

Table 1. True or-gate values

| Input wire $\alpha$ | Input wire $\beta$ | Output wire $\gamma$ |
|---|---|---|
| 0 | 0 | 0 |
| 0 | 1 | 1 |
| 1 | 0 | 1 |
| 1 | 1 | 1 |

Table 2. Or-gate GCs

| Input wire $\alpha$ | Input wire $\beta$ | Output wire $\gamma$ | Garbled value |
|---|---|---|---|
| $K_\alpha^0$ | $K_\beta^0$ | $K_\gamma^0$ | $\mathrm{E}^s_{K_\alpha^0 K_\beta^0}(K_\gamma^0)$ |
| $K_\alpha^0$ | $K_\beta^1$ | $K_\gamma^1$ | $\mathrm{E}^s_{K_\alpha^0 K_\beta^1}(K_\gamma^1)$ |
| $K_\alpha^1$ | $K_\beta^0$ | $K_\gamma^1$ | $\mathrm{E}^s_{K_\alpha^1 K_\beta^0}(K_\gamma^1)$ |
| $K_\alpha^1$ | $K_\beta^1$ | $K_\gamma^1$ | $\mathrm{E}^s_{K_\alpha^1 K_\beta^1}(K_\gamma^1)$ |

The encrypted values in Tables 1 and 2 shows the true value of each or-gate wire. The output wire $\gamma$ can only output 0 when $\alpha$ and $\beta$ both input 0. Otherwise, output wire $\gamma$ outputs a 1. According to Table 2, each item in the scheme corresponds to a combination of the input wire values and the garbled output value encryption. The resulting lookup table is the GC [39]. Because keys are generated randomly by $P_\alpha$, $P_\beta$ cannot be used to deduce the true $P_\alpha$ value. When the encrypted result is determined, we can obtain the true values without any true input value details, in a process like that in Figure. 1 (b).

Another challenge is how to avoid $P_\alpha$ knowing the true value of $P_\beta$ by recording the transfer information. This problem is resolved by using the oblivious transfer (OT) technique, an essential protocol where a sender transfers one of the many pieces of information to a receiver but remains oblivious as to which piece has been accepted [40]. This technology, which involves two parties, was invented by Kilian Joe [41]. The OT processes are as follows:
- Input: Sender inputs a pair of strings ($K_\beta^0$, $K_\beta^1$), and the receiver inputs a bit (0 or 1).
- Output: Only the receiver outputs $K_\beta^\sigma$, $\sigma \in \{0,1\}$.

This one-out-of-two OT can efficiently prevent the data from leaking.

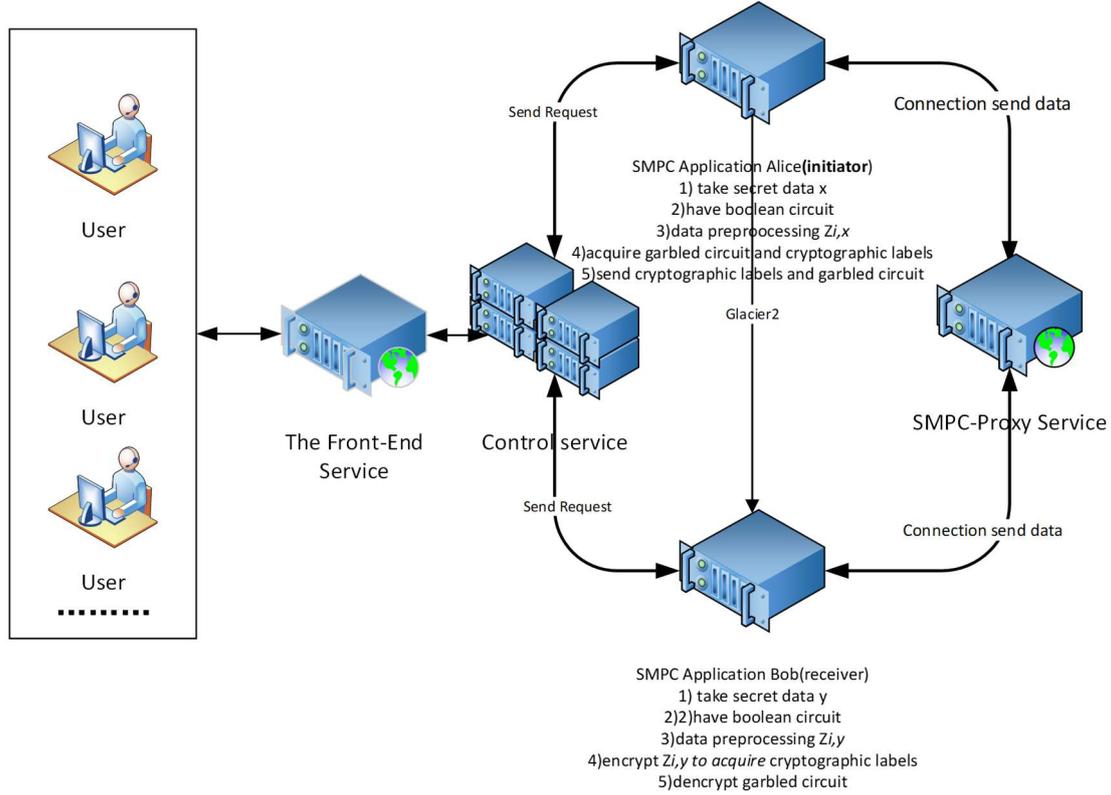

**Figure 2.** This figure illustrates how users can utilise the SMPC service to obtain results without acquiring secret data.

**Table 3.** Functions of the five services

| Service | Function |
| --- | --- |
| Front-end | Send the request to the control server, provide a visual user interface for users, and show the outcome. |
| Control | A middleware and a distributor. When the frond-end server transmits the parameters of the user requests, this server can select the address of the network interface from the metadata table, which records the location where data is stored, and the node information table, which includes the service uniform resource locator (URL). Then, it analyses the data and assigns tasks to Alice and Bob. When the calculation is completed, the control server obtains the result and returns it to the users. |
| SMPC-proxy | A medium between Alice and Bob before they build the connection. It not only can be installed with any other service but can also deploy on a server independently. |
| Alice | The Initiator has the secret data, the number of taxi companies in the Jiangsu Province, and a Boolean circuit. The column names include "gid" (the serial number of each city) and the company number. |
| Bob | The receiver has the evening light data in the Jiangsu Province and a Boolean circuit. The column names include gid (the serial number of each city) and the light. |

### 3.2 SMPC constructions

The system has five components, including the front-end service, SMPC application Alice (initiator), SMPC application Bob (receiver), SMPC-proxy service, and control service, as shown in Figure. 2. These services are deployed across four servers, where the front-end server has the front-end service, the control service is run in the control server, and the SMPC applications are deployed in the Alice and Bob servers, storing different sensitive data. Additionally, the SMPC-proxy service can operate on any server. In this study, we install it together with Alice. The functions of each service are shown in Table 3.

We configured the four servers to meet the experimental requirements. All machines have a quad-core core processing unit, with 32 GB of memory and a 500-GB hard disk, and run using a CentOS7.5 operating system.

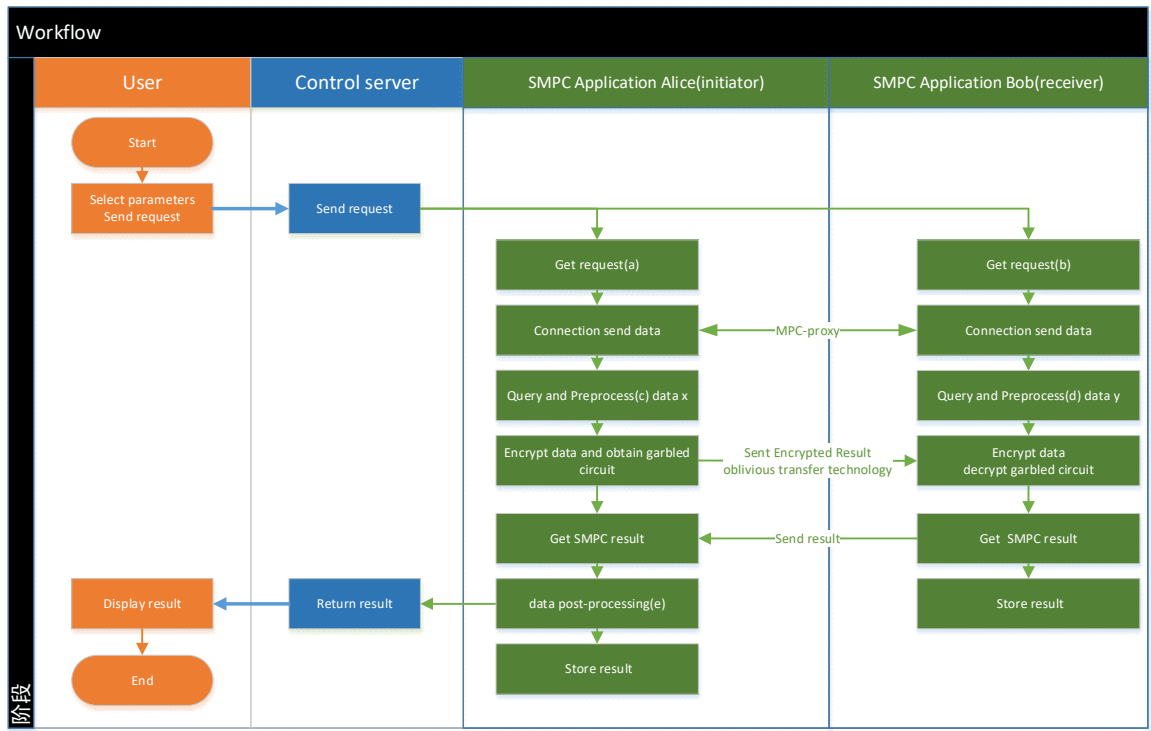

**Figure 3.** The workflow of the entire system.

Figure 3 illustrates how the entire system works. There are four servers in the flow chart, including one for the user (with front-end service), a control server (with control service), an Alice server (with SMPC-proxy and an SMPC initiator), and a Bob server (with an SMPC receiver).

The workflow begins with users selecting parameters on the website according to their requirements. Then, the front-end service sends the request to the control server, which can obtain the service URL using the metadata table and can successively assign tasks to Bob and Alice. Bob (receiver) is first and prepares for the requests. Second is Alice (initiator), who sends the connecting request to the SMPC-proxy, and then the request goes to Bob. If he agrees, he tells the proxy, and Alice is informed. Then, Bob and Alice build the connection.

The following steps are for the calculation. Both Alice and Bob must complete data pre-processing, such as selecting data from the databases and making the necessary calculations. Alice creates six strings and a GC to encrypt her private data as illustrated in Section 2.1. Bob then receives those keys, decrypting the results, and shares the results with Alice. The users then obtain the results from Alice.

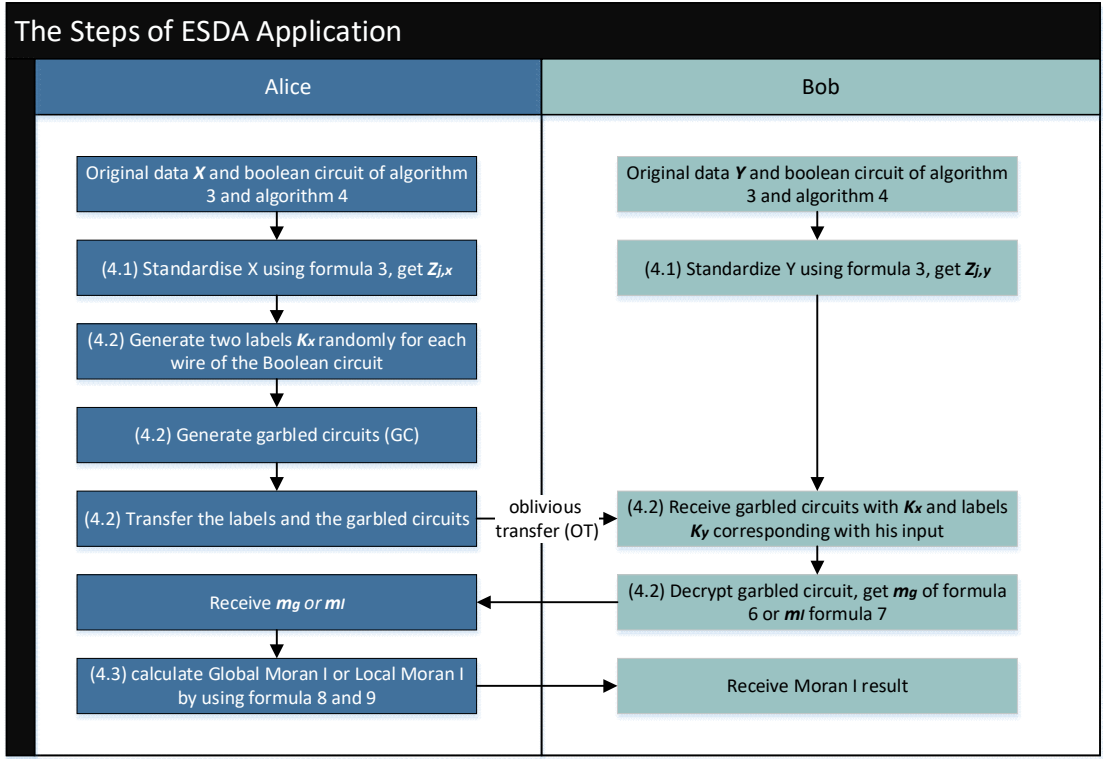

**Figure 4.** ESDA application steps

## 4. The application of ESDA

If we used real private data, we could not publish the original data and experimental results. Therefore, data was collected from a public website instead as an example to demonstrate the protocol. The data included the number of taxi companies in the Jiangsu Province and evening light data of the Jiangsu Province. We assume that the two participants cannot obtain the data of the other participant from internet.

We also selected the ESDA algorithm as an example to exhibit our method. In this section, we demonstrate how to build ESDA protocols in SMPC like those in Figure 4. The process is divided into three components: preprocessing, SMPC-processing, and post-processing.

### 4.1 Moran's I and local indicators of spatial association (LISA)

Global Moran' I is the test for spatial correlation. This test statistic is formulated as a quadratic in terms of the variables that are being tested for spatial correlation [42, 43]. The formula is as follows:

$$I = \frac{\sum_i(\sum_j w_{i,j} z_{j,y} z_{i,x})}{\sum_i z_{i,x}^2} \qquad (1)$$

When I is greater than 0, then the data have a positive spatial correlation. I = 0 would indicate no correlation. If I is less than 0, the data have a negative spatial correlation. Global measurements can aptly describe the general features of the special data. However, it hides the local pattern and cannot be used for mapping Therefore, to reveal the single regional spatial correlation, Anselin [44] created LISA for local Moran's I, which can be used to detect positive and negative autocorrelation hotspots [45]. The formula as follows:

$$I_i = \frac{(n-1) \times z_{i,x} \times \sum_j w_{i,j} z_{j,y}}{\sum_i z_{i,x}^2} \qquad (2)$$

where the original data standardises the variables by subtracting the sample mean and then dividing by the standard deviation (SD). These two functions are similar, and their common indicators are as follows:

$$z_{i,x} = \frac{x_i - \bar{x}}{\sigma_x} \qquad (3)$$

$$z_{i,y} = \frac{y_i - \bar{y}}{\sigma_y} \qquad (4)$$

In addition, SD can be calculated as:

$$\sigma = \sqrt{\frac{1}{N}\sum_{i=1}^{N}(a_i - \bar{a})^2} \qquad (5)$$

Both global Moran's I and local Moran's I are popular. Their results can be visualised in a scatter plot, where the I value is the slope of the line [46]. In this study, we introduce the two-party security calculation framework to obtain these two indicators, without exposing any original data. The steps include pre-processing, SMPC-processing, and post-processing.

*4.2 Pre-processing*

In the first step, we preprocess x Alice, who must standardise x as formula 3, and Bob does the same with y.

| Algorithm 1. X standardisation |
| --- |
| Data: the number of taxi companies (X) and night light (Y) in the Jiangsu Province |
| Result: $z_{j,x}$ or $z_{j,y}$ |
| 1. **function** std (array K) |
| 2.     **for** j→size(K) do |
| 3.         Sum+=(K-average(K))×(K-average(K)) |
| 4.     **end for** |
| 5. **return** (Sum/Size(K)) 1/2 |
| 6. |
| 7. **function** main() |
| 8.     **for** i→size(X) do |
| 9.         $z_{j,x}$[i] =[x-average(X)]/std(X) |
| 10.     **end for** |
| 11. **end** function |

*4.3 SMPC-processing*

We use Frutta, a language of the SMPC framework, to write the algorithms. Because the efficiency of this framework is not particularly high, we must minimise the calculation process. Therefore, we only use SMPC to calculate the interactive part.

In Moran's I $m_g = \sum_i(\sum_j w_{i,j} z_{i,x} z_{i,y})$ (6)

In LISA $m_l = z_{i,x} \times \sum_j w_{i,j} z_{j,y}$ (7)

where wi,j is the normalised inverse distance weight matrix. The pseudocode is algorithms 3 and 4. Then, we must compile the algorithm to obtain Boolean circuit file information like that in Table 4.

Table 4. Compilation information of the two algorithms

| Item | Global Moran's I | Local Moran's I |
| --- | --- | --- |
| Compile time | 0.155936 s | 0.163897 s |
| numGates | 145334 | 150233 |
| numWires | 146166 | 151065 |
| input1 size | 416 bit | 416 bit |
| input2 size | 416 bit | 416 bit |
| output1 size | 32 bit | 416 bit |
| NOT | 1740 | 1740 |
| AND | 39508 | 40579 |
| OR | 104086 | 107914 |
| Zip time | 0.075979 s | 0.074589 s |
| Size | 1.33 MB | 1.37 MB |

Table 4 shows that the local Moran's I is slightly more complicated than the global Moran's I. The number of gates (numGates) and wires (numWires) increase by nearly 5000 from the global Moran's I to the local one, meaning that the AND-gate and OR-gate both rise. This is because the local Moran's I has to output each I value, while the global Moran's I only outputs the value once.

Next, Alice generates two labels randomly for each wire of the Boolean circuit. Then, she encrypts each output wire label with two labels of the input wires of each gate to obtain the GCs [30]. Afterward, she sends the labels corresponding to her input and the GCs to Bob. OT technology prevents data leakage during the data transmission. When Bob acquires the labels corresponding to his original input, he begins to decrypt the GC, but he does not know $z_{i,x}$. Bob must also encrypt his data $z_{i,y}$. Therefore, Bob decrypts the GC by using labels from both himself and Alice. Then, the result m is obtained.

**Algorithm 2.** SMPC result of global Moran's I

Data：the weight of each city $w_{i,j}$, $z_{j,y}$ and $z_{j,x}$
Result：$m_g$

12. **function** global_moral ($w_{i,j}$, $z_{j,x}$, $z_{j,y}$)
13.     **for** i→size($z_{j,y}$) do
14.         sum←0;
15.         **for** j→size($z_{j,x}$) do
16.             sum ←sum+ w[i][j]* $z_{j,x}$[j];
17.         **end for**
18.         output ← output + sum* $z_{j,y}$[i];
19.     **end for**
20. **return** output
21. 
22. **function** main()
23.     $m_g$ ← global_moran ($w_{i,j}$, $z_{j,x}$, $z_{j,y}$)
24. **end function**

**Algorithm 3.** SMPC result of local Moran's I

Data：the weight of each city $w_{i,j}$, $z_{j,y}$ and $z_{j,x}$
Result：$m_l$

1. **function** local_moran ($w_{i,j}$, $z_{j,x}$, $z_{j,y}$)
2.     **for** i→size($z_{j,x}$) do
3.         sum←0;
4.         **for** j→size($z_{j,y}$) do
5.             sum ←sum+ w[i][j]* $z_{j,x}$[j];
6.         **end for**
7.         output[i] ←sum* $z_{j,y}$[i];
8.     **end for**
9. **return** output
10. 
11. **function** main()

12.     $m_l \leftarrow$ local_moran $(w_{i,j}, z_{j,x}, z_{j,y})$
13.  **end function**

*4.4 Post-processing*

To obtain the global and local Moran's I, Alice must accomplish data post-processing. The formulas and algorithms are as follows:

Global Moran's I: $\quad I = \frac{m_g}{\sum_i z_{i,x}^2} \quad$ (8)

Local Moran's I: $\quad I_i = \frac{(n-1) \times m_l}{\sum_i z_{i,x}^2} \quad$ (9)

**Algorithm 4.** Calculating Moran's I result

Data：$m_g, z_{j,x}$
Result： I
1.  **function** square (z)
2.      **for** i→size(z) **do**
3.          sum←0;
4.          **for** j→size(z) **do**
5.              sum ←sum+ $z_{i,j} \times z_{j,i}$
6.          **end for**
7.          output ← output +sum
8.      **end for**
9.  **return** output
10. 
11. **function** main()
12.     $I = \frac{m_g}{\text{square}(z_{j,x})}$
13. **end function**

**Algorithm 5.** Calculating Moran's I result

Data：$m_g, z_{j,x}$
Result： I
1.  **function** square (z)
2.      **for** i→size(z) **do**
3.          sum←0;
4.          **for** j→size(z) **do**
5.              sum ←sum+ $z_{i,j} \times z_{j,i}$
6.          **end for**
7.          output ← output +sum
8.      **end for**
9.  **return** output
10. 
11. **function** main()
12.     $I = \frac{(size(z_{j,x})-1) \times m_l}{\text{square}(z_{j,x})}$
13. **end function**

Alice knows p. Therefore, she only needs to calculate $\sum_i z_{i,x}^2$. Depending on the above steps, she can obtain the global and local Moran's I without exposing the original data.

The next step is hypothesis testing. We shuffle the data randomly and input it again, repeating the process K times, to determine $I_h = (I_1, I_2, I_3, \ldots\ldots, I_K)$. The number of $I_h$ greater than and less than the original I are recorded by using greater-I and lesser-I, respectively. So we can calculate p using the following equation:

$$p = \frac{R + 1}{K + 1} \qquad (10)$$

| Algorithm 6. Calculating $p$ |
|---|
| Data：$I_h = (I_1, I_2, I_3, \ldots\ldots, I_K), K$ |
| Result：presult[K] |
| 1.  **function** main() |
| 2.      for i→K do |
| 3.          If $I_i$> i |
| 4.            presult[i] ← presult[i] + 1; |
| 5.          End if |
| 6.           output ← output +sum |
| 7.      **end for** |
| 8.  **end function** |

where R is equal to the minimum of the greater-I and lesser-I. The SD is another target, which represents stability:

$$SD = (I_1 - \bar{I}_h)^2 + (I_2 - \bar{I}_h)^2 + \cdots + (I_k - \bar{I}_h)^2 \qquad (12)$$

| Algorithm 7. Calculating SD |
|---|
| Data：$I_h$ |
| Result：SD |
| 1.  **function** main() |
| 2.      SD = 0 |
| 3.      **for** i→size($I_h$) do |
| 4.          SD ← SD + $(I_i - $avertage$(I_h))^2$ |
| 5.      **end for** |
| 6.  **end function** |

## 5. Results

By completing the steps outlined in Section 4, we can obtain the result of global Moran's I, 0.29468779, which is extremely close to the result from other data-sharing platforms, 0.294873, indicating that night light and the number of taxi companies have a positive correlation.

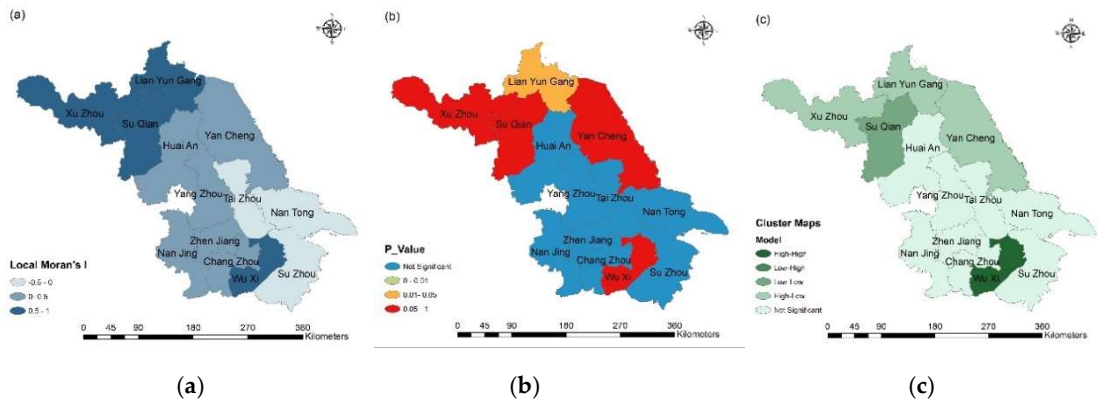

(a)　　　　　　　　　　(b)　　　　　　　　　　(c)

**Figure 5.** results of LISA. (**a**) the distribution of Local Moran I in Jiangsu Province; (**b**) P-value of each cities of Jiangsu Province; (**c**) Cluster maps of Jiangsu Province

Additionally, Figure 5 displays the LISA results. There is a clear relationship (P-value is less than 0.05) between the two datasets. By analysing the local Moran's I value of the light data in 2010 and the number of taxi passenger companies in 2008, it can be seen that the local Moran's I in Taizhou, Nantong, and Suzhou is less than 0, exhibiting a negative correlation; the local Moran's I in Nanjing, Zhenjiang, and Changzhou were between 0 and 0.5, displaying a positive correlation.

local Moran's I values in Xuzhou, Suqian, Lianyungang, and other cities were greater than 0.5, indicating that the light values have a large positive correlation with the number of passenger transport companies in the adjacent regions.

We compare our results with those of the other platforms and calculate the accuracy to ensure reliability. The accuracy is calculated by the following equation.

$$P = X - \hat{X} \qquad (13)$$

where $X$ is the SMPC result, and we see the other platforms result as the true value ($\hat{X}$). Based on Table 5, we know that most of absolute errors are no greater than 0.008, which indicates reliability and meets the requirements of many studies. In contrast, the time efficiency was insufficient. Moran's I is a simple calculation; therefore, it only takes 20 s. However, LISA is more complex; hence, the process takes approximately 35 s. This time estimate is based on a low internet delay of 0.4 ms. When the internet delay increases to 40 ms, the time consumption increases to over 10 min (Table 6).

**Table 5.** SMPC accuracy

| City | SMPC | Other platforms | Accuracy |
|---|---|---|---|
| Nanjing | 0.1971360077188468 | 0.1979123 | 0.000776 |
| Lianyungang | 0.5727442254838897 | 0.5765237 | 0.003779 |
| Huaian | 0.29020802877024726 | 0.2908422 | 0.000634 |
| Yancheng | 0.3248602163616162 | 0.3266383 | 0.001778 |
| Changzhou | 0.1624733933688088 | 0.1645943 | 0.002121 |
| Taizhou | -0.02500597625869832 | -0.0285191 | -0.00351 |
| Suzhou | -0.41837787872054255 | -0.4188669 | -0.00049 |
| Suqian | 0.8986661832641364 | 0.9061707 | 0.007505 |
| Nantong | -0.02116243400970703 | -0.0233131 | -0.00215 |
| Xuzhou | 0.6289066428863808 | 0.6329164 | 0.00401 |
| Wuxi | 0.6559079614057656 | 0.6595777 | 0.00367 |
| Zhenjiang | 0.19294680428045138 | 0.1962196 | 0.003273 |
| Yangzhou | 0.05673981170691771 | 0.0577783 | 0.001038 |
| Moran's I | 0.29468779 | 0.294873 | 0.000185 |

**Table 6.** Time efficiency

| | Internet delay | Moran's I | LISA |
|---|---|---|---|
| Other platforms | 0.4 ms | <1 s | <1 s |
| SMPC | 0.4 ms | 20 s | 35 s |
| Other platforms | 40 ms | <1 s | <1 s |
| SMPC | 40 ms | 10 min | 30 min |

## 6. Case studies

One potential application scenarios is considered, and our approach is contrasted with alternatives. In the medical area, when a virus such as human immunodeficiency virus or coronavirus starts to spread, the government and researchers want to determine its origin and trend. However, both the disease data, which is stored in the hospital system, and personal information of citizens, stored by the government, are private. Many patients are reluctant to share their private data because they may risk exposing their physical condition, which they do not wish to disclose. In this case, government and hospitals can build an SMPC system, and then they can acquire the spatial autocorrelation and local Moran's I at smaller scales and analyse the relationship between the disease and location without exposing any sensitive data. Furthermore, if income must be taken into account, then banks can also be incorporated into the framework. Using this method, patients do not need to worry about their data security, and research can be performed while conforming to data security policies.

## 7. Discussion and Conclusions

The main contribution of this study is filling the gap between SMPC and spatial data sharing. Users and providers can both access the spatial analysis results without going through a trusted third party. There are several potential advantages of this research. The main benefit of this data sharing security approach is that two participators can protect their private data from leaking, compared with the traditional method, by guaranteeing accuracy and meeting demands. Secondly, in our examination, we considered the characteristics of the algorithms and spatial data; hence, our result can be visualised without compromising any privacy. However, time efficiency is a weakness of the proposed platform compared to that of other platforms; although we attempted to improve the algorithms, data, and code structures, users must still wait for several minutes.

As the next research step, we will implement our findings to solve practical problems. Future research can also improve the performance and algorithm efficiency. Finally, we plan to provide an improved visualisation tool to help users customise their results.

**Funding:** This research was funded by National Key R&D Program of Chian, grant number 2016YFB0502200 and 2016YFB0502204. The work also was funded by Joint Fund for Space Science and Technology, grant number 4201420100041. In addition, we get the support from the Fundamental Research Funds for the Central Universities, as well as Special Research Fund of State Key Laboratory of Information Engineering in Surveying, Mapping and Remote Sensing (LIESMARS)

**Author Contributions:** All authors made a substantial contribution to this manuscript. Shen Shuo write the original manuscript as well as write the program. Zhu Xinyan and An Rui were in charge of the study design, also reviewed the paper. Ma Yanlei collected the field data and visualized the result, at the same time, he improved the structure of manuscript. Xie Xiang, Sun Lilin and Xie Hongjungave many help in the designation of the model and application of the model. All authors read and approved the final manuscript.

**Conflicts of Interest:** The authors declare no conflict of interest.